\documentclass[twocolumn]{aastex631}

\usepackage{graphicx,times}             
\usepackage{natbib}
\usepackage{indentfirst}
\usepackage{amssymb,amsmath}
\usepackage{multirow}
\usepackage{threeparttable}

\bibpunct{(}{)}{;}{a}{}{,}

\shorttitle{Galaxy evolution in Abell 2142}
\shortauthors{Qu ChengGong et al.}
\graphicspath{{./}{figures/}}

\begin{document}

\title{The Way to Quench: Galaxy Evolution in A2142}

\correspondingauthor{Heng Yu}
\email{yuheng@bnu.edu.cn}

\author{ChengGong Qu}
\affiliation{Department of Astronomy, Beijing Normal University, Beijing 100875, China}

\author{Heng Yu}
\affiliation{Department of Astronomy, Beijing Normal University, Beijing 100875, China}

\author{Antonaldo Diaferio}
\affiliation{Dipartimento di Fisica, Università di Torino, Via P. Giuria 1, I-10125 Torino, Italy}
\affiliation{Istituto Nazionale di Fisica Nucleare (INFN), Sezione di Torino, Via P. Giuria 1, I-10125 Torino, Italy}

\author{Jubee Sohn}
\affiliation{Astronomy Program, Department of Physics and Astronomy, Seoul National University, 1 Gwanak-ro, Gwanak-gu, Seoul 08826,
Republic of Korea}
\affiliation{Smithsonian Astrophysical Observatory, 60 Garden Street, Cambridge, MA 02138, USA}

\author{DengQi Liu}
\affiliation{Department of Astronomy, Beijing Normal University, Beijing 100875, China}

\begin{abstract}

 We show how the star formation activity of galaxies is progressively inhibited from the outer region to the center of the massive cluster A2142. From an extended spectroscopic redshift survey of 2239 galaxies covering a circular area of radius $\sim 11$~Mpc from the cluster center, we extract a sample of 333 galaxies with known stellar mass, star formation rate, and spectral index $D_n4000$.
We use the Blooming Tree algorithm to identify the substructures of the cluster and separate the galaxy sample into substructure galaxies, halo galaxies and outskirt galaxies. The substructure and halo galaxies are cluster members, whereas the outskirt galaxies are only weakly gravitationally bound to the cluster. For the cluster members, the star formation rate per stellar mass decreases with decreasing distance $R$ from the cluster center. Similarly, the spectral index $D_n4000$ increases with $R$, indicating an increasing average age of the stellar population in galaxies closer to the cluster center. In addition, star formation in substructure galaxies is generally more active than in halo galaxies and less active than in outskirt galaxies, proving that substructures tend to slow down the transition between field galaxies and cluster galaxies.
We finally show that most actively star forming galaxies are within the cluster infall region, whereas most galaxies in the central region are quiescent.

\end{abstract}


\section{Introduction} \label{sec1}

According to the current model of the formation of cosmic structures, clusters of galaxies form by gravitational instability from perturbations in the initial matter density field. Small groups of galaxies flow along the filaments of the cosmic web and contribute to the formation and evolution of galaxy clusters. In hierarchical scenarios, increasingly massive clusters form, on average, at increasingly later times \citep{2007MNRAS.381.1450N,2009MNRAS.398.1150B}.

Spectrophotometric properties of galaxies are correlated with the density of the galaxy environment. For example,
galaxies in the local universe show two distinct distributions in the
color-magnitude diagram: a red sequence, mostly due to 
early-type 
galaxies,
and a blue cloud, mostly due to star-forming late-type  galaxies \citep{2001AJ....122.1861S,2003ApJ...594..186B}. Galaxies on the red sequence are generally located in the dense central regions of
galaxy clusters, whereas blue-cloud galaxies populate  less dense environments\citep{1980ApJ...236..351D,1984ApJ...281...95P,2004ApJ...615L.101B,2013MNRAS.433.2667R,2018ApJ...852..142C,2019MNRAS.487.5572M}.
In the current model of galaxy formation, late-type galaxies might evolve into early-type galaxies through various processes, including galaxy merging, tidal stripping, and ram pressure stripping \citep{1999ApJ...510L..15B,2004MNRAS.348..811T}.
While falling from the outskirts to the center of a massive galaxy cluster,
galaxies are likely to be affected by these types of interaction.
Although the shock of the hot intracluster medium (ICM) acting on the cold gas of the galaxy can sometimes increase the star formation activity
\citep{2019MNRAS.486L..26S},  this ram pressure stripping mostly removes cold gas from the galaxy
and inhibits star formation \citep{1972ApJ...176....1G,2000ApJ...540..113B,2013A&A...557A.103J,2015Natur.521..192P,2020A&A...638A.126D,2020AJ....159..218T}.
The timescale associated with this starvation mechanism is  $\sim 4$~Gyr \citep{2015Natur.521..192P},
generally longer than the timescale for ram pressure stripping of $\sim 0.5- 4$Gyr.

 In a cluster, the local galaxy density is correlated with the distance from the cluster center, and
the fraction of late-type galaxies increases with clustrocentric distance, as happens, for example, in the Perseus cluster \citep{2020A&A...640A..30M}.
Similarly, in A2029, the spectral index $D_{n}4000$ of the cluster galaxies, an indicator of the age of their stellar population,  decreases with clustrocentric distance, suggesting younger stellar populations in the outer galaxies, as expected for late-type galaxies \citep{2019ApJ...872..192S}.

Most rich clusters exhibit some amount of substructure in the galaxy distribution \citep{1982PASP...94..421G,2013MNRAS.436..275W}.
Since galaxies in substructures have relative velocities comparable to the velocity dispersion of stars in galaxies, the probability of galaxy mergers increases \citep{2015A&A...579A...4G,2016A&A...586A..63Z}.
Indeed, many early-type galaxies are in substructures \citep{2014A&A...562A..87E}, suggesting that galaxy mergers might have already taken place before the galaxies were completely accreted by the cluster.
Because of the diversity of environments within a galaxy cluster  and its outer region, investigating the properties of cluster galaxies provides crucial information on galaxy evolution.

A2142 is a massive galaxy cluster at redshift z $\sim$ 0.0898.
It is located at the center of a supercluster  connected to the large-scale filamentary structure \citep{2020A&A...641A.172E}.
The
cold fronts of A2142 observed in X-rays are probably the result of a sloshing cool core in the central region \citep{2001ApJ...562L.153M,2005ApJ...618..227T,2011ApJ...741..122O}.
A galaxy group  that is undergoing ram pressure stripping is also observed near the radius  $ R_{500}$ \citep{2014A&A...570A.119E}.
The outskirts of the cluster
are dominated by star-forming blue galaxies, unlike the inner region \citep{2018A&A...620A.149E}.
Although the dense environment of the central region of the cluster has an impact on the evolution of galaxies,
many galaxies are within high-density substructures flowing toward the cluster along  filaments that surround it. The relation between galaxy properties, clustrocentric distance, and local environment is thus complicated by the presence of substructures.

The caustic method based on a hierarchical clustering algorithm \citep{2015Yu} can be used
to identify the substructures of clusters. The algorithm was successfully applied to A85 \citep{2016Yu} and A2142
\citep{2018ApJ...863..102L}. The Blooming Tree algorithm is an updated version of the algorithm \citep{2018ApJ...860..118Y}. Here, we plan to apply the Blooming Tree algorithm to identify the substructures of A2142 and constrain the relation between galaxy properties and local environment.

This paper is organized as follows. In Section \ref{sec2}, we present our data. In Section \ref{sec3}, we separate our sample into three subsamples according to their membership of the cluster, substructures, or outer region. We distinguish star-forming galaxies from quiescent galaxies,  and discuss the relation between  their physical properties and their environment.
We summarize our results in Section \ref{sec4}.
Throughout this paper, we adopt a Wilkinson Microwave Anisotropy Probe standard cosmological model with $\Omega_m$ = 0.272, $\Omega_{\Lambda }$ = 0.728, and $H_0 = 70.4\, km\, s^{-1}\, Mpc^{-1}$ \citep{2011ApJS..192...18K}. All the errors we mention are $1\sigma$.

\section{Observational data} \label{sec2}

\begin{figure}[htbp]
	\centering
	\includegraphics[width=1\linewidth]{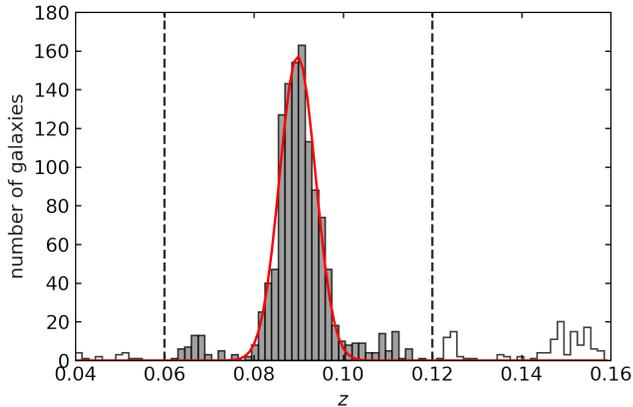}
	\caption{Redshift distribution of the galaxies in the field of view of A2142. The hollow bars show the distribution
		of the $z$-available galaxies. The gray bars show the distribution of the $z$-slice galaxies, the 1186 galaxies in our sample whose redshift is in the range
		$[0.06, 0.12]$. The vertical dashed lines indicate this redshift range. The red solid line is the Gaussian fit to the distribution of the $z$-slice galaxies after 3$\sigma$ clipping.}
	\label{fig1}
\end{figure}

\citet{2018ApJ...863..102L} compiled a spectroscopic redshift survey of 2239 galaxies in the field of view of A2142. Hereafter, we call these 2239 galaxies the $z$-available galaxies. This catalog covers a circle of radius 0.$^\circ$56 from the cluster center, whose celestial coordinates are R.A. = 239.$^\circ$5833 and decl. = 27.$^\circ$2334. This angular radius corresponds to a radius of 10.8 Mpc at the cluster redshift $z=0.09$. Figure \ref{fig1} shows the redshift distribution of the galaxies around this redshift. For the analysis of the structure of A2142, we consider the 1186 galaxies with redshift in the range [0.06,0.12]. Hereafter, we call these 1186 galaxies the $z$-slice galaxies. The redshift distribution of the $z$-slice galaxies is shown by the gray bars in Fig.~\ref{fig1}. The red solid line is the Gaussian fit to this distribution obtained after 3$\sigma$ clipping.

The star formation rates(SFRs) and the stellar masses, $M_\star$, of the $z$-slice galaxies are collected from the GALEX–SDSS–WISE Legacy Catalog \citep[GSWLC,][]{2018ApJ...859...11S}. This catalog is based on photometric data in multiple bands, including UV data taken by the Galaxy Evolution Explorer \citep[GALEX,][]{2005ApJ...619L...1M} and optical data taken by the Sloan Digital Sky Survey \citep[SDSS,][]{2009ApJS..182..543A}.

We also consider the spectral index $D_{n}4000$, the ratio of the average flux densities in the narrow continuum bands 3850-3950 \AA\ and 4000-4100 \AA\ \citep{Balogh_1999}.  This spectral index correlates  with the age of the stellar population that contributes most of the electromagnetic emission in the optical band \citep{bruzual1983spectral, poggianti1997indicators}. The  $D_{n}4000$ values are queried from the database of SDSS \citep{2003ApJ...599..971H}.

\begin{figure}[htbp]
	\centering
	\includegraphics[width=1\linewidth]{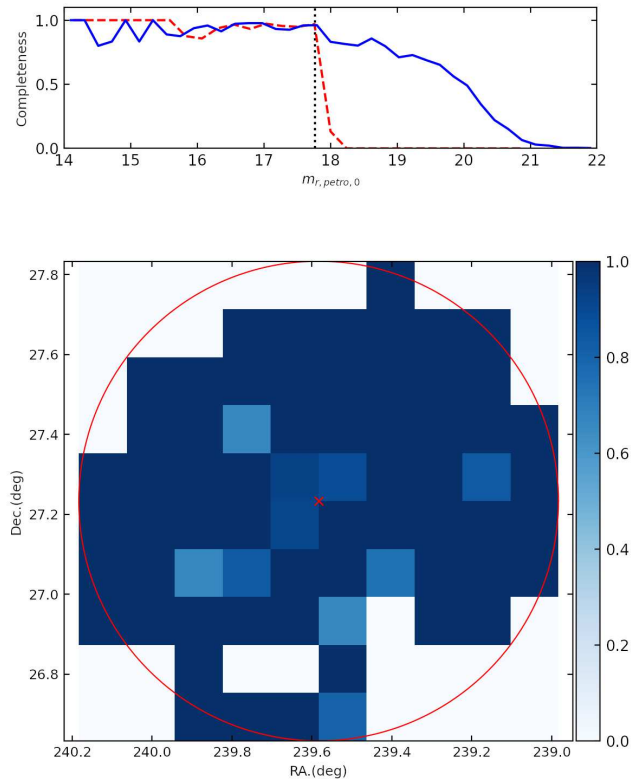}
	\caption{Top panel: the spectroscopic completeness of the $z$-available galaxies as a function of the Petrosian $r$-band magnitude (blue line).
     The red dashed line shows the ratio between the number of data-available galaxies and the number of $z$-slice galaxies.
     Bottom panel:  the distribution of the ratio between the number of data-available galaxies and the number of $z$-slice galaxies on the sky. In this panel only galaxies with $m_{r,Petro,0} < 17.77$ are considered.
     The red circle shows a circle of radius $0.^\circ$56  around the cluster center indicated by the red cross.}
	\label{fig2}
\end{figure}

There are 333 galaxies, out of the 1186 $z$-slice galaxies, with  SFRs, stellar mass $M_\star$, and $D_{n}4000$ available. Hereafter, we call these 333 galaxies the data-available galaxies. The spectroscopic completeness of the 2239 $z$-available galaxies as a function of the Petrosian $r$-band magnitude is shown by the blue line in the top panel of Fig. \ref{fig2}.
The decrease in spectroscopic completeness at magnitudes fainter than  18 mag is caused by the Petrosian $r$-band magnitude limit $m_{r,Petro,0} < 17.77$ of the spectroscopic galaxy sample of  SDSS \citep{Balogh_1999}. The red dashed line indicates the ratio between the 333 data-available galaxies and the 1186 $z$-slice galaxies.

The bottom panel of Fig. \ref{fig2} shows the spatial distribution of the ratio between the number of data-available galaxies and the number of $z$-slice galaxies. We limit the computation of this ratio to the galaxies with  $m_{r,Petro,0} < 17.77$. We have  303 data-available galaxies and 319 $z$-slice galaxies brighter than this magnitude limit.
 The two-dimensional map shown in Fig. \ref{fig2} has $10 \times 10$ pixels for a squared field of view of $1.^\circ$12 on a side.
 The overall ratio is $303/319=0.95$ with a standard deviation  $0.286$. In the panel, the pixels outside the red circle contain no data.

\section{Data analysis} \label{sec3}

In  Sect. \ref{subsec1} we use the Blooming Tree algorithm to split our galaxy sample into three subsamples: halo, substructure, and outskirt galaxies. In Sect. \ref{subsec2} we distinguish star-forming galaxies from quiescent galaxies;
Sect. \ref{sfrdn} discusses the relation between the star-formation rate per unit mass (the specificSFR, or
sSFR) and the spectral index $D_n4000$; Sect. \ref{radial} focuses on the radial distribution of sSFR and $D_n4000$, and Sect. \ref{pps} discusses the galaxy distribution in the $R$$-$$v$ diagram.

\subsection{Halo, substructure, and outskirt galaxies} \label{subsec1}

The Blooming Tree algorithm is a method for identifying substructure based on the hierarchical clustering algorithm \citep{2018ApJ...860..118Y}. It arranges all the galaxies in the field of view into a tree, or dendrogram. The arrangement is based on the pairwise  projected binding energy, which is estimated from the location  of the galaxies on the sky and from their redshift \citep{1999MNRAS.309..610D}. By adopting a proper density contrast parameter $\Delta \eta$, we can trim the tree into distinct structures: $\Delta\eta$ is the difference between two values of $\eta$, the former associated with the structure and the latter associated with the surrounding background structure; $\eta$ combines the line-of-sight velocity dispersion, the size, and the number of galaxies in the structure; increasing values of $\Delta\eta$ identify increasingly dense structures \citep[see][for further details]{2018ApJ...860..118Y}.

We apply  the Blooming Tree algorithm to the sample of 1186 $z$-slice galaxies. By setting $\Delta \eta =5$, we identify 684 cluster galaxies. By increasing the density contrast to $\Delta \eta = 25$, we identify denser structures: we find 16 structures with more than five member galaxies. The basic properties of these 16 structures are listed in Table \ref{tab:substructure}:
$n_{\mathrm{g}}$ is the number of member galaxies, $n_{\mathrm{d}}$ is the number of member galaxies with known SFR, $M_\star$, and $D_{n}$4000, $z_{\mathrm{sub}}$ is the average redshift of the structure, and $v_{\mathrm{disp}}$ is the velocity dispersion of the structure. All the 480 members of the 16 structures, which are named sub1 to sub16, belong to the 684 cluster galaxies identified with the contrast parameter $\Delta\eta=5$. We associate the remaining 204 out of these 684 galaxies with a diffuse component indicated by grp0 in Table \ref{tab:substructure}.

\begin{table}
\centering
\begin{center}
\caption{\label{tab:substructure} Properties of the galaxy structures. $n_{g}$ is the number of member galaxies, and $n_{d}$ is the number of member galaxies with known SFR, $M_{\star}$,and $D_{n}4000$. $z_{sub}$ is the average redshift of the structure and $v_{disp}$ is its velocity dispersion.}

\begin{tabular}{|l|llll|}
  \hline\hline
  GroupID & $n_{\mathrm{g}}$ & $n_{\mathrm{d}}$ & $z_{\mathrm{sub}}$ & $v_{\mathrm{disp}}(\mathrm{km}\, \mathrm{s}^{-1})$  \\
  \hline
  cluster & 684 & 189 & 0.0898 & 912$\pm$11\\
  \hline
  grp0 & 204 & 66 & 0.0895& 1059$\pm$10\\
  sub1 & 178 & 43 & 0.0902& 786$\pm$10\\
  sub2 & 81 & 26 & 0.0884 & 477$\pm$10\\
  sub3 & 41 & 10 & 0.0929 & 464$\pm$10\\
  sub4 & 26 & 7 & 0.0870  & 303$\pm$10\\
  sub5 & 22 & 3 & 0.0916  & 403$\pm$11\\
  sub6 & 18 & 5 & 0.0888  & 266$\pm$10\\
  sub7 & 17 & 8 & 0.0895  & 356$\pm$7\\
  sub8 & 17 & 4 & 0.0960  & 318$\pm$14\\
  sub9 & 12 & 1 & 0.0858  & 353$\pm$15\\
  sub10 & 12 & 1 & 0.0892 & 283$\pm$11\\
  sub11 & 12 & 2 & 0.0870 & 351$\pm$6\\
  sub12 & 11 & 4 & 0.0897 & 334$\pm$6\\
  sub13 & 10 & 3 & 0.0907 & 219$\pm$12\\
  sub14 & 9 & 2 & 0.0946  & 304$\pm$10\\
  sub15 & 7 & 4 & 0.0906  & 202$\pm$8\\
  sub16 & 7 & 0 & 0.0887  & 347$\pm$6\\

\hline

\end{tabular}
\end{center}
\end{table}

\begin{figure*}[htbp]
     \centering
    \includegraphics[width=1\linewidth]{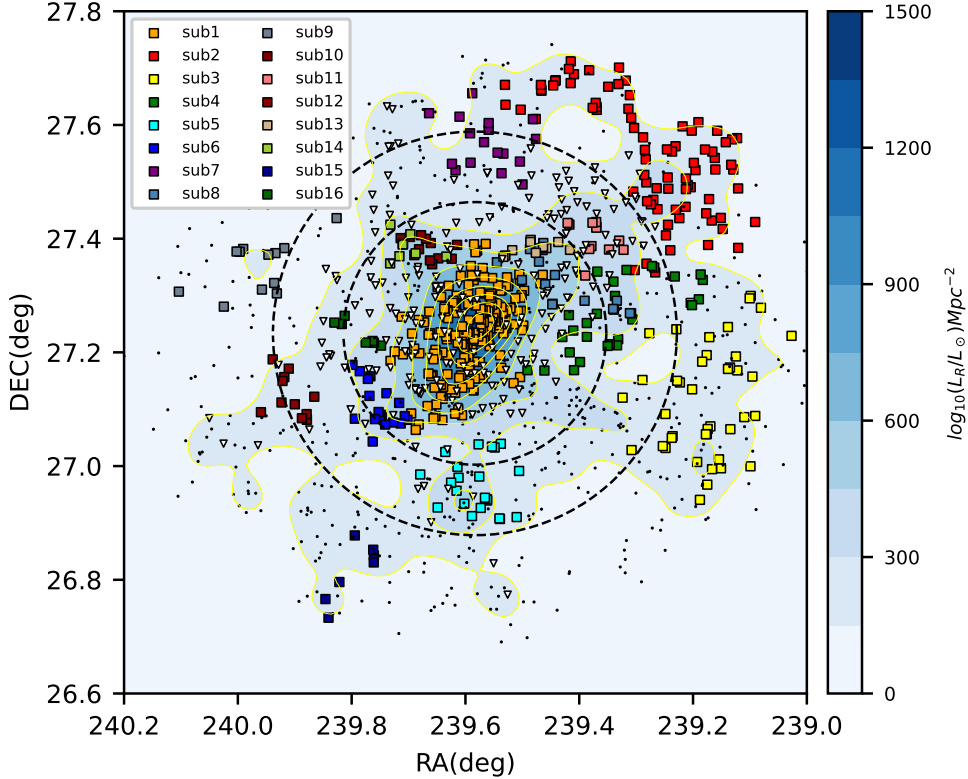}
     \caption{Distribution on the sky of the 1186 $z$-slice galaxies superimposed on the distribution of their Petrosian $r$-band luminosity. The galaxy sample separates into 382 halo galaxies and 302 substructure galaxies, totalling to 684 cluster members, and 502 outskirt galaxies.
     The open triangles show the members of grp0, the orange squares show the members of the core sub1, the other colored squares show the members of the structures from  sub2 to sub16, and the black dots show the outskirt galaxies. The two black dashed circles have radius $R_{500} = 1.408$ Mpc and $R_{200}= 2.160$ Mpc at the cluster redshift $z=0.09$ \citep{tchernin2016xmm}.
     }
     \label{fig4}
 \end{figure*}

The distribution of the $z$-slice galaxies on the sky is shown in Fig. \ref{fig4}, where the 480  members of the structures
are represented as colored squares; the 204 cluster members associated with grp0 are represented as open triangles; the remaining 502 $z$-slice galaxies which belong neither to the cluster nor to any structure, are represented by black dots.

The galaxies are plotted on top of the map of Petrosian $r$-band luminosity density.
The density map is computed from the 1186 $z$-slice galaxies by assuming that the  $r$-band luminosity $L_R$ of each galaxy is smoothed with a 2D Gaussian window of $2^\prime$ width.
\citep[see][for details]{2013MNRAS.436..275W}.
Figure~\ref{fig4} shows that the distribution of the luminosity density is generally consistent with the distribution of the structures, as expected.

The structures from sub2 to sub16 are distinct components that we identify as cluster substructures. The structure sub1 (orange squares in Fig.~\ref{fig4}) is located at the cluster center and we identify this structure with the cluster core.
 We define halo galaxies to be the galaxies in the core and in the structure grp0, substructure galaxies to be members of the structures from sub2 to sub16, and outskirt galaxies to be the $z$-slice galaxies that are neither halo nor substructure galaxies.

The 333 data-available galaxies, with known SFR, $M_\star$, and $D_n$4000, separate into 109 halo galaxies (grp0 and sub1), 80  substructure galaxies (sub2 to sub16), and 144 outskirt galaxies. In the following analysis and discussion we  focus on these three galaxy subsamples.

\subsection{Star-forming and quiescent galaxies} \label{subsec2}

\begin{figure}[htbp]
	\centering
	\includegraphics[width=1\linewidth]{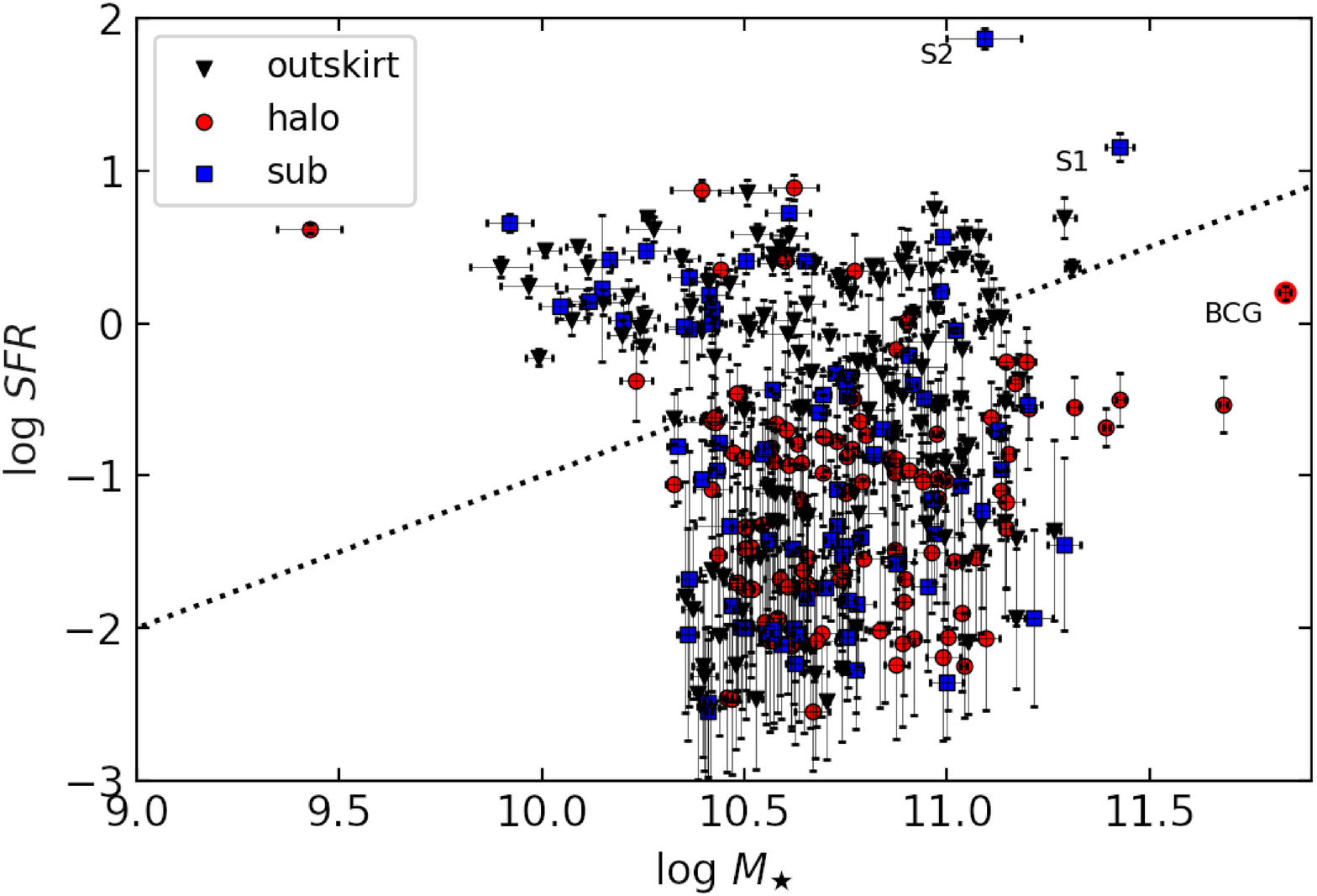}
	\caption{The distribution of the 333 data-available galaxies in the SFR$-M_{\star}$ plane. The red circles, blue squares, and black triangles represent halo, substructure, and outskirt galaxies, respectively. The black dotted line shows the specific SFR, sSFR = $10^{-11}$~yr$^{-1}$, separating SF galaxies from quiescent galaxies.
    S1 and S2 are the brightest galaxies of substructures sub14 and sub3, respectively. BCG is the brightest galaxy of A2142.}
\label{fig5}
\end{figure}

Cluster galaxies at low redshift generally distributed into two distinct groups in the plane of stellar mass versus star formation rate, $M_{\star} -$SFR \citep{noeske2007star,2015Natur.521..192P}: the two groups distinguish the star-forming (SF) galaxies, with smaller values of the spectral index $D_{n}4000$, and the quiescent galaxies,  with larger values of $D_{n}4000$. The 333 data-available galaxies in our sample show this bimodal distribution, with most galaxies being massive and quiescent  (Fig. \ref{fig5}). It is worth noting that there are only star-forming galaxies when $M_{\star} < 10^{10.4} M_{\odot} $. To keep the completeness of our sample, we do not remove them. However, our later results remain the same without these less massive starforming galaxies.

We consider the sSFR, defined as the SFR per unit stellar mass $M_\star$. We separate the SF from the quiescent galaxies according to the threshold sSFR = $10^{-11}$~yr$^{-1}$  \citep{2011McGee,2012Wetzel}. The 333 data-available galaxies separate into 93 SF galaxies and 240 quiescent galaxies. Table \ref{tab:2} lists how these galaxies are distributed into halo, substructure, and outskirt galaxy samples.
The fraction of SF galaxies steadily increases from the halo sample to the outskirt sample. This trend suggests that star formation is progressively quenched from the outskirt galaxies to the substructure and the halo galaxies. The larger fraction of SF galaxies in the substructures than in the halo sample is also consistent with the scenario where the substructure galaxies entered the cluster more recently than the halo galaxies, and star formation in substructure galaxies is less inhibited than in halo galaxies.

There are two SF galaxies with  SFR larger than $10$~$M_{\odot }$yr$^{-1}$. We label these galaxies S1 and S2. S1, with $\log [{\mathrm{SFR}/({\mathrm{M}}_\odot{\mathrm{yr}}^{-1})]} = 1.866$, is the brightest galaxy of the substructure sub14; S2, with $\log [{\mathrm{SFR}/({\mathrm{M}}_\odot{\mathrm{yr}}^{-1})]} = 1.156$, is the brightest galaxy of the substructure sub3.
Sub14 is a substructure falling toward the center of the cluster at a high speed, as shown by \citet{2014A&A...570A.119E,2017A&A...605A..25E}, and \citet{2018ApJ...863..102L}. Ram pressure stripping might enhance the star formation rate of S1 \citep{roberts2021lotss}.  The image of S1 is also disturbed, suggesting an ongoing galaxy merger \citep{2018ApJ...863..102L}. In contrast, the large SFR of S2 might derive from its nature of grand-design spiral galaxy, whose face-on image appears undisturbed.

\begin{table}
\small
\centering
\caption{\label{tab:2} Star-forming and Quiescent Data-available Galaxies.$n_{d}$ is the number galaxies with known SFR, $M_{\star}$, and $D_{n}4000$.}

\begin{tabular}{llll}
   \hline\hline
   Galaxy Sample   & $n_{\mathrm{d}}$ & SF & Quiescent\\
   \hline
   Total & 333 & 93 (27.9\%) & 240 (72.1\%)\\
   \quad Halo & 109 & 10 (9.2\%) & 99 (90.8\%)\\
   \quad Substructure & 80 & 20 (25.0\%) & 60 (75.0\%)\\
   \quad Outskirt & 144 & 63 (43.8\%) & 81 (56.3\%)\\
   \hline
\end{tabular}
\end{table}

\begin{figure}[htbp]
	\centering
	\includegraphics[width=1\linewidth]{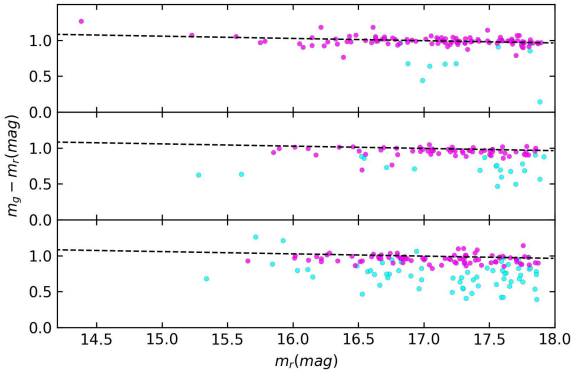}
	\caption{The color–magnitude ($m_{g}-m_{r}$)$-m_{r}$ diagram for the halo galaxies (top), substructure galaxies (middle), and  outskirt galaxies (bottom). The magenta and cyan dots show the quiescent and SF galaxies, respectively. The dashed line in each pane is the red-sequence fit to the quiescent halo galaxies in the top panel: $m_{g} - m_{r}$ = $-0.0314m_r$ + $1.528$. }

\label{fig3}
\end{figure}

The decreasing fraction of quiescent galaxies from the halo to the outskirt sample is also apparent in the
decreasing fraction of galaxies lying on the red-sequence relation in the color-magnitude diagram (Fig. \ref{fig3}).
The mean colors of the quiescent galaxies are comparable in the three samples:  0.99 $\pm$ 0.07,  0.95 $\pm$ 0.06, and 0.96 $\pm$ 0.06 for the halo, substructure, and outskirt samples, respectively.

\subsection{The sSFR$-D_{n}4000$ relation}
\label{sfrdn}

\begin{figure}[htbp]
     \centering
     \includegraphics[width=1\linewidth]{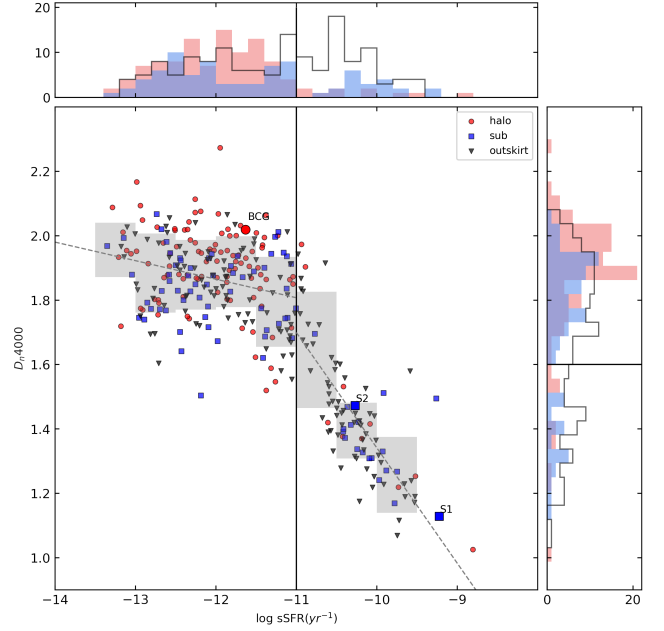}
     \caption{ Top panel: the distribution of sSFR   for the halo galaxies (red histogram), the substructure galaxies (blue histogram), and the outskirt galaxies (hollow histogram). Right panel: same as the top panel for $D_n4000$. Bottom left panel: the relation between sSFR and $D_{n}4000$. The two dotted lines are linear fits to the SF and quiescent galaxies separately. For illustrative purposes, the entire galaxy sample is separated into bins of fixed width on the $\log({\mathrm{sSFR/yr}^{-1}})$ axis: for each of these bins, the shaded areas show the rms values of $D_{n}4000$ around the mean.  BCG, S1, and S2 are the central bright galaxies of A2142, sub14, and sub3, respectively.}
     \label{fig6}
\end{figure}

Figure \ref{fig6} shows the distributions of the spectral index $D_{n}4000$ and the sSFR for the galaxies in our three samples. This figure also shows the correlation between these two quantities.
For the halo and substructures galaxies, the distributions peak at small sSFR and large $D_{n}4000$, whereas for the outskirt galaxies the distributions appear somewhat flat. This different behavior indicates a correlation between the environment and the galaxy properties.

Our galaxy sample indeed confirms the expected anticorrelation between sSFR and $D_n4000$ \citep{kauffmann2004environmental}:
$D_{n}4000$ increases with the age of the stellar population \citep{Balogh_1999} and is thus expected to increase with decreasing sSFR if sSFR decreases with increasing age of the stellar population \citep{duarte2022mass}.

We separately fit the sSFR$-D_n4000$ relation for the SF galaxies and the quiescent galaxies, and find $D_{n}4000 = -0.39\log({\mathrm{sSFR/yr^{-1}}}) - 2.24$ and $D_{n}4000 = -0.06\log({\mathrm{sSFR/yr^{-1}}}) + 1.17$ for the two samples, respectively.
The relation is steeper for the sample of SF galaxies than for the quiescent galaxies, suggesting that the star formation activity  gradually decreases in increasingly denser environments.

\subsection{The radial distribution of sSFR and $D_n4000$}
\label{radial}

\begin{figure}
    \centering
     \includegraphics[width=1\linewidth]{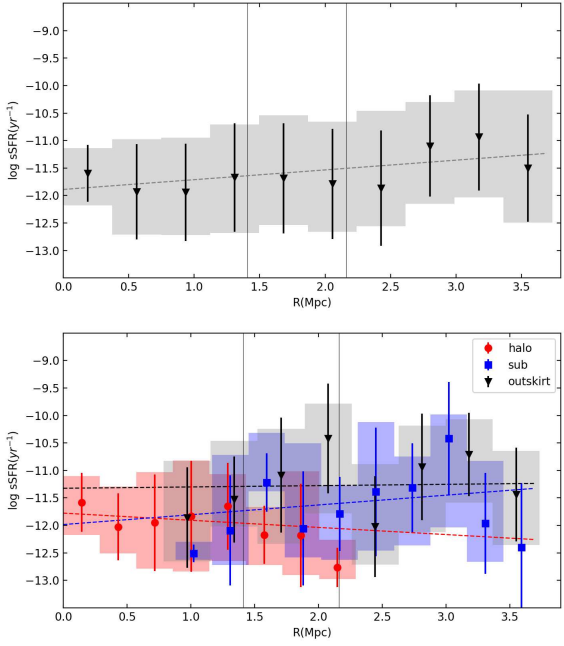}
     \caption{The dependence of sSFR on the projected clustrocentric radius $R$. Top panel: the entire galaxy sample with the rms values of sSFR around the mean in each of the 10 evenly spaced radial bins (shaded area). The dots indicate the median values of each bin. The dotted line shows the best linear fit.
     The two vertical solid lines show the two radii $R_{500}=1.408$~Mpc and $R_{200}=2.16$~Mpc. Bottom panel: same as the top panel for the three galaxy subsamples separately: halo (red), substructure (blue), and outskirt galaxies (black/gray).
     }
     \label{fig8}
\end{figure}

The properties of cluster galaxies are closely related to the local galaxy density, which, in turn, generally depends on the clustrocentric distance  \citep{odekon2018effect,coccato2020formation,2020A&A...640A..30M}. Figure \ref{fig8} shows the dependence of the sSFR on the projected radius $R$ from the cluster center for the entire galaxy sample (top panel) and for the three galaxy samples separately (bottom panel).
The data are divided into 10 equally spaced radial bins. The median value of each bin is given. Their rms values are shown with shading.
Despite the large scatter, our entire sample shows that sSFR increases with $R$. The substructure galaxies are mainly responsible for this trend. Indeed, the outskirt galaxies show a flat relation, with larger sSFRs than substructure galaxies, on average,  and the halo galaxies show a slightly decreasing relation.

\begin{figure}
    \centering
     \includegraphics[width=1\linewidth]{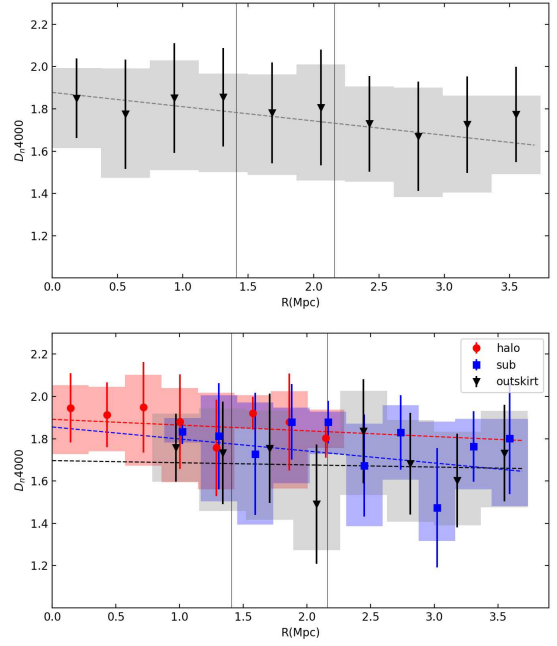}
     \caption{Same as Fig. \ref{fig8} for the $D_n4000$$-$$R$ relation. }

     \label{fig9}
\end{figure}

Figure \ref{fig9} shows the dependence of $D_n4000$ on $R$ \citep{Balogh_1999}. It mirrors the dependence of sSFR on $R$, because of the correlation between sSFR and $D_n4000$ shown in Fig. \ref{fig6}:
for the entire sample, $D_n4000$ decreases with increasing $R$, similarly to the galaxies in A2029 \citep{2019ApJ...872..192S}. As in the sSFR-$R$ relation, this trend is mostly due to the substructure galaxies, whereas halo and outskirt galaxies have shallower relations, with the values of $D_n4000$ of the outskirt galaxies smaller, on average, than those for the other two galaxy samples.

Figures \ref{fig8} and \ref{fig9} show that the average values of sSFR and $D_n4000$ of the substructure galaxies in the cluster center are comparable to the values of the halo galaxies. Similarly, at large radii, these quantities of the substructure galaxies are comparable to the values of the outskirt galaxies.
This result suggests (1) that the substructure galaxies are sensitive to the environment of their own substructure, and (2) that substructures tend to slow down the transition from field galaxies to cluster galaxies. This scenario is consistent with results of simulations, which suggest that orphan galaxies that have lost their subhalos are more vulnerable to environmental effects than those that still have them\citep{2018Cora}. Those orphan galaxies belong to the diffuse halo galaxies here.

\begin{figure}
    \centering
     \includegraphics[width=1\linewidth]{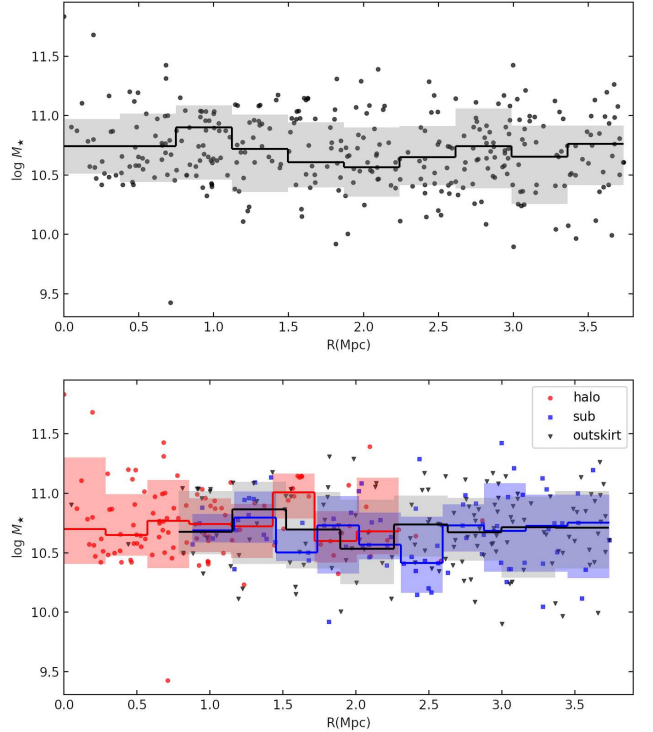}
     \caption{Same as Fig. \ref{fig8} for the log$M_{\star}$$-$$R$ relation. The solid black line indicates the median value in each bin.}

     \label{rm}
\end{figure}

Considering the sSFR depends strongly on the stellar mass, the mass segregation effect could bias our result. We check the distribution of radial stellar mass log$M_{\star}$ of the three subsamples and find their median masses are consistent at all radii, as Fig. \ref{rm} shows.

%

\subsection{The $R$$-$$v$ diagram} \label{pps}

We know only three out of the six phase-space coordinates of each galaxy in the field of view of A2142: the two celestial coordinates and the line-of-sight velocity. This limited knowledge prevents us from grasping the full dynamics of the cluster and its structure.
Nevertheless, from the distribution of galaxies in the $R$$-$$v$ diagram, namely the line-of-sight velocity versus the projected distance from the cluster center, we can infer the global depth of the gravitational potential well of the cluster, or, equivalently, the escape velocity from the cluster \citep{1997ApJ...481..633D,1999MNRAS.309..610D,2011MNRAS.412..800S}, and identify the galaxies that are members of the cluster \citep{2013ApJ...768..116S}. This information can be extracted for a large interval of projected distances from the cluster center, from the central region to radii much larger than the virial radius, in regions where the dynamical equilibrium hypothesis does not hold and where the galaxies surrounding the cluster are falling into it for the first time.

\begin{figure}
     \centering
      \includegraphics[width=1\linewidth]{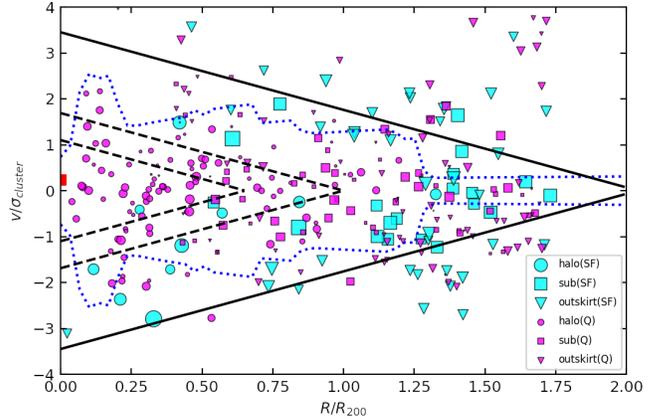}
      \caption{The $R$$-$$v$ diagram of the 333 data-available galaxies of A2142. The data-available galaxies consist of 80 substructure galaxies (squares), 109 halo galaxies (circles), and 144 outskirt galaxies (triangles). Cyan and magenta symbols show SF and quiescent galaxies, respectively.
      The symbol size is proportional to the specific SFR.
      The red square on the left is the BCG. The blue dotted lines show the caustic location.
      The black solid and dashed lines are described in the text.}

      \label{fig10}
 \end{figure}

Figure \ref{fig10} shows the $R$$-$$v$ diagram, or projected phase-space (PPS) diagram, of our three galaxy samples.
The blue dotted lines show the location of the caustics derived in \citet{2020ApJ...891..129S}. The caustics are related to the escape velocity from the cluster \citep{1997ApJ...481..633D,1999MNRAS.309..610D}. According to \citet{2013ApJ...768..116S}, the sample of galaxies within the caustics contains $(95\pm 3)$\% of the real members and   is contaminated by $\sim 8$\% of interlopers within 3$R_{200}$. The caustic technique thus represents a valid procedure to identify cluster members in real data.

Alternatively, \citet{{2016MNRAS.463.3083O}} adopt an approximate relation to identify cluster members. Their relation is based on dark matter-only  simulations. By defining as interlopers those satellite dark matter halos with distance, in real space, $r_{3d} > 2.5 r_{vir}$, with $r_{vir}$ the cluster virial radius, they find that, in the $R$$-$$v$ diagram, the line $v/\sigma_{3d} = -(4/3) R/r_{vir} +2$  roughly separates the region of the $R$$-$$v$ diagram dominated by the cluster members from the region dominated by interlopers.
The black solid lines in Fig. \ref{fig10} show the line of \citet{{2016MNRAS.463.3083O}}, where we set $\sigma_{3d}=\sqrt{3}\sigma_{cluster}$ and  $R_{200}/r_{vir}=0.73$.
The black solid lines are roughly consistent with the caustic location and thus appear to be a reasonable proxy for the caustics. For the sake of simplicity, we adopt these black solid lines, rather than the caustics, as the cluster boundaries.

Figure \ref{fig10} shows two additional sets of  black dashed lines: they have the same slope as the black solid lines and cross the $v/\sigma_{cluster}=0$ axis at $R_{200}$ and $R_{500}$, respectively. We adopt these lines  in the $R$$-$$v$ diagram as the counterparts of $R_{200}$ and $R_{500}$ in real space.

The distribution of our galaxy samples in the $R$$-$$v$ diagram is generally consistent with the identification of the cluster members derived in Sect. \ref{subsec1}:
most outskirt galaxies (triangles), which are not expected to be cluster members, lie outside the regions identified by the caustics or the black solid lines, whereas most halo and substructure galaxies, which are expected to be cluster members, are within these regions.

We now investigate the star formation activities of the galaxies in the infall region of the cluster. We define the infall region as the band of the $R$$-$$v$ diagram between the black solid line and the  black dashed line crossing the point $(R_{200},0)$. We consider the specific SFR, sSFR, and the spectral index $D_n4000$ as a function of the distance $\Delta d$ of each galaxy from the black solid line: $\Delta d$ is thus the segment perpendicular to the black solid line joining the galaxy and the black solid line.
According to the analysis of the orbits of galaxies falling into clusters in numerical simulations \citep[e.g.,][]{2017ApJ...838...81Y,2019MNRAS.484.3968A}, a galaxy that is falling into the cluster traces a trajectory roughly parallel to the black solid line in the $R$$-$$v$ diagram; the radial coordinate $R$ of  this trajectory clearly  decreases during the galaxy infall. Therefore, larger $\Delta d$ implies larger initial radial distance of the falling galaxy.

\begin{figure}[htbp]
     \centering
      \includegraphics[width=1\linewidth]{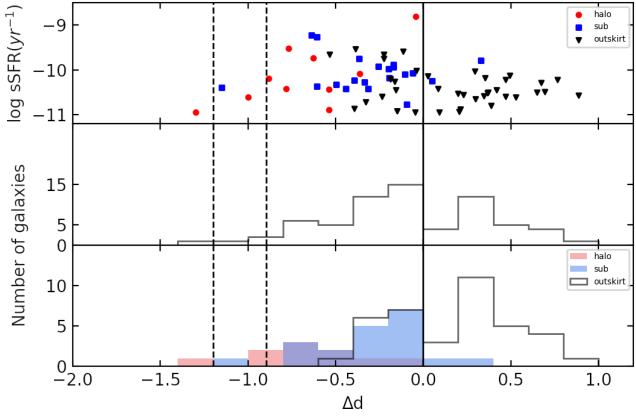}
      \caption{Top: relation between sSFR and $\Delta d$ for the 93 SF galaxies in the data-available galaxy sample. Middle: the distribution of $\Delta d$ for the full sample of 93 SF galaxies. Bottom: the distribution of $\Delta d$ for the halo, substructure, and outskirt SF galaxies separately.
      The two vertical black dotted lines indicate $R_{200}$ ($\Delta d$ = $-0.9$) and $R_{500}$ ($\Delta d$ = $-1.2$), respectively. The vertical black solid line is the boundary line  $\Delta d=0$.}
      \label{fig11}
 \end{figure}

Figure \ref{fig11} shows the $\Delta d$ distribution of the 93 SF galaxies, namely the data-available galaxies with $\log(\mathrm{sSFR/yr}^{-1})> -11$. Out of these 93 galaxies, 30 are cluster members: specifically there are 20  substructure galaxies and 10 halo  galaxies. Out of these 20 and 10  galaxies, 17 and 8, respectively, lie in the infall region, namely in the band between the black solid line and the outer black dashed line in the $R$$-$$v$ diagram.
Therefore, 83\% (25 out of 30) of the cluster members that are actively forming stars are in the infall region. Our sample thus demonstrates, as expected, that the dense intracluster medium within $R_{200}$ inhibits the star formation activity.
The only SF galaxy (LEDA 1801474) within $R_{500}$ is a halo galaxy. The SDSS image of this galaxy suggests that its star formation activity is triggered by an ongoing merger.

\begin{figure}[htbp]
     \centering
      \includegraphics[width=1\linewidth]{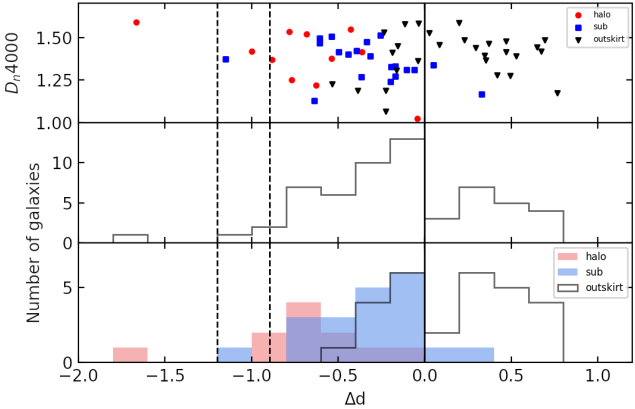}
      \caption{Same as Fig. \ref{fig11} for the 82 blue galaxies with $D_{n}4000 < 1.6$.
      }
      \label{fig12}
 \end{figure}

Figure \ref{fig12} shows the $\Delta d$ distribution of the 82 galaxies with spectral index $D_n4000 < 1.6$. We call these galaxies blue galaxies. Their distribution is similar to the distribution of sSFR in Fig. \ref{fig11}.
There are 26 (84\%) out of 31 blue galaxies that are cluster members, namely either substructure or halo galaxies, in the infall region.
There are 17 out of 20 substructure blue galaxies, and 9 out of 11 halo blue galaxies.
The only blue galaxy within $R_{500}$ is a halo galaxy (SDSS J155827.26+271300.3). Its color might be contaminated by a nearby blue object, which is only 4 arcsec away.

Figures \ref{fig11} and \ref{fig12} show that the SF and blue galaxies that are cluster members are concentrated in the infall region, namely  in the PPS region located between the black dashed line corresponding to $R_{200}$ and the black solid line. This result indicates that the dense ICM environment substantially inhibits
 the star formation activity of the galaxies once they enter the region within $R_{200}$. In addition, the transition from star forming galaxies to quiescent galaxies substantially ends at radii larger than $R_{500}$.

\section{Summary} \label{sec4}

We compiled a catalog of 333 galaxies from a spectroscopic redshift survey of 2239 galaxies in the field of view of the cluster A2142 \citep{2018ApJ...863..102L}. The survey covers a circular area of radius $\sim 11$~Mpc from the cluster center. Each of the 333 galaxies has measured stellar mass $M_\star$, SFR, and spectral index $D_{n}$4000. We use the Blooming Tree algorithm, an algorithm for the identification of cluster substructure \citep{2018ApJ...860..118Y}, to separate our sample into  three subsamples: the halo, the substructure, and the outskirt galaxies. The halo and the substructure galaxies are cluster members. The outskirt galaxies are still in the outer region of the cluster, but, according to the Blooming Tree algorithm, their gravitational bond to the cluster is weak.

We investigate the relation between the environment and the star formation activity of the galaxies in these three subsamples. Our main conclusions are as follows.

\begin{itemize}
     \item
     The specific SFR, sSFR=SFR/$M_\star$, is larger in the outskirt galaxies and smaller in the halo galaxies. In addition, the sSFR increases, on average, with increasing distance from the cluster center; similarly, the spectral index $D_{n}4000$, which is an indicator of the age of the stellar population,  decreases with increasing distance from the cluster center. Both results show that the star formation activity tends to be inhibited a high density environment.
     \item The sSFR of substructure galaxies is intermediate between the sSFR of halo galaxies and  that of outskirt galaxies; in addition, the sSFR depends on the environment of the substructure of the galaxy, being smaller, on average, for galaxies in substructures that are close to the cluster center, and larger for galaxies in substructures that are in the outer region of the cluster. The spectral index $D_n4000$ shows the same behavior. This result demonstrates that substructures tend to slow down the transition between field galaxies and cluster galaxies. \item Galaxies that are actively forming stars  mostly lie in the cluster infall region, roughly between $R_{200}$ and the turn-around radius: star formation is progressively inhibited while approaching $R_{200}$ and substantially quenched within $R_{200}$.

\end{itemize}

Our analysis demonstrates the relevance of spectroscopic redshifts for investigating the connection between the physical properties of galaxies and their environment. For this goal, our Blooming Tree algorithm proves efficient at associating the galaxies with the composite structures of a cluster. The application of the Blooming Tree algorithm to data from future extensive spectroscopic surveys, such as Euclid \citep{2022Euclid} or LSST \citep{2019LSST}, is thus expected to greatly enhance our comprehension of galaxy evolution in clusters.

\section*{Acknowledgments}
We thank the referee sincerely for his/her valuable comments and suggestions in the report.
This work was supported by Bureau of International Cooperation, Chinese Academy of Sciences GJHZ1864.
A.D. acknowledges partial support from the INFN grant InDark.


\bibliography{bibtex}
\bibliographystyle{aasjournal}

\end{document}